
{\bf 1. Introduction}
\vskip 1truecm

The influence of lattice discreteness on the properties of
nonlinear systems having kink solutions was investigated by
several authors [1]-[10]. These studies have pointed out a large
variety of effects, including modification of kink velocity and
its form and leading sometimes to the pinning of the kink on the
lattice. But on the other side numerical investigations [9] have revealed the
existence of narrow solitary waves propagating in certain
discrete lattices without energy loss. There is also a close
connection of these problems with the complete integrability of
specific models in nonlinear cotinuum and discrete systems [11]-[13].

Since the pioneering paper of  Krumhansl and Schrieffer [14] the
role played by the elementary excitations of kink type in the
thermodynamics of one-dimensional nonlinear systems was
investigated by many authors ([15]-[20] and the references
therein). As concerns the influence of lattice discreteness on
these thermodynamic properties this has received a smaller
attention [20]-[23]. The class of 1-D systems mostly discussed
is described by the Hamiltonian (in the notations of CKBT [15])
$$H~=~\sum_{i} lA\left({1\over2}\dot\phi_i^2~+~{c_0^2\over 2l^2}
(\phi_{i+1}-\phi_i)^2~+~\omega_0^2 V(\phi_i)\right)\eqno(1)$$
where the nonlinearity enters only through the potential
$V(\phi)$, assumed to have at least two degenerate minima with
nonvanishing curvature. It is the merit of Trullinger and Sasaki
[21] to have shown that the first lattice corrections are taken
into account if the potential $V(\phi)$ is replaced by an
effective potential
$$V_{eff}~=~V(\phi)~-~{l^2\over24d^2}\left({dV\over d\phi}\right)^2
{}~+~O\left({l^4\over d^4}\right),\eqno(2)$$
where $d={c_0\over \omega_0}$ is the mean width of the static
kink. In the present paper we intend to complete their results
and determine also the lattice corrections to the multi-kink
contributions to the free energy.

The partition function and the free energy are usually
calculated using the transfer integral operator (TIO) method. In
the thermodynamic limit the lowest eigenvalues of TIO play the
most important role. In the displacive/continuum limit,
${l\over d}<<1$, it is possible to transform the Fredholm
integral equation representing the TIO into a differential
equation. This is easily done when the two-body interaction in
(1) has an harmonic character \footnote{*}{As was shown by Guyer and Miller
[17] even for more complicated interactions between nearest
neighbours it is possible to write a formal expression for
the partition function and the free energy using a cumulant
expansion.} [17],[19]. One obtains
$$exp(-\gamma V(\phi))exp\left({\gamma\over2m^\ast}D^2\right)
\Phi_n (\phi)~=~exp(-\gamma\tilde\epsilon_n)\Phi(\phi).\eqno(3)$$
Here $\gamma=\beta lA\omega^2_0,~~~~~~m^\ast=(\beta
Ac_0\omega_0)^2,~\Phi_n$ is the eigenvector of TIO corresponding
to the eigenvalue $\epsilon_n$ and
$$\tilde\epsilon_n~=~\epsilon_n~+~{1\over\gamma}~ln~{l\over d}
\sqrt{{2\pi\over\gamma}}.\eqno(3a)$$
The left hand side of (3) contains the product $e^A\cdot e^B$
with A and B two noncommuting operators, and in order to obtain
the claimed differential equation we have to put it as $e^C$,
where the operator $C$ can be written in terms of $A$ and $B$
using the Baker-Hausdorff formula [20],[21],[22]. This formula
turns to be a mixed expansion in the small parameters ${l^2
\over d^2}$ and ${\gamma\over m^\ast}={l\over
d}(m^\ast)^{-{1\over2}}$ and consequently is very convenient for
the proposed pourposes. The first lattice correction is obtained
if one retaines only the terms proportional with ${l^2\over
d^2}$. Contrary the opinion of Trullinger and Sasaki [21] the
same result is found if one work with the symmetrised or
non-symmetric form of TIO. It is possible to show through a set
of transformations and neglecting systematically higher powers
in ${l^2\over d^2}$, that the eigenvalues of TIO, with the first
lattice corrections included, can be found from the equation [21]
\footnote{**}{The discrepancy between the results of [21] and
[20] comes from an inadequate treatment of a certain equation
obtained in an intermediate step of the proof in our previous
paper [20].}
$${1\over 2m^\ast}~{d^2\Psi_n\over d\phi^2}~+~(\tilde\epsilon_n-
V_{eff}(\phi))\Psi_n~=~0.\eqno(4)$$
The result is very general and valid for any kind of potential
function $V(\phi)$.

In the next section the leading terms as an asymptotic expansion
of the lowest eigenvalues of (4) will be found in the low
temperature limit, $m^\ast>>1$. The method used, briefly
sketched in the Appendix, allow us to make a clear distinction
between the contributions of phonons and of the various kink
sectors. Explicit expressions for the lattice corrections to the
kink and kink-kink contributions to the free energy will be
given. The last section is devoted to a discussion of the
results in the spirit of the CKBT [15] phenomenology of
independent phonons and renormalised kinks. As it is known [15],
[16], [19] this phenomenology is exact in the low temperature
limit of the continuum models and the results of Trullinger and
Sasaki [21] and those of the present paper strongly support the
conjecture that it remains valid even when the first lattice
corrections are taken into account.
\vskip 2truecm

{\bf 2. Lattice Corrections to the Free Energy}
\vskip 1truecm
   Using (3a) the free energy per unit length becomes
$$F~=~{1\over \beta l}~ln~\beta h\omega_{0}{d\over l}~+~A\omega^{2}
_{0}\tilde\epsilon_{0}\eqno(5)$$
where $\tilde\epsilon_{0}$ is the lowest eigenvalue of eq. (4).

In the low temperature limit $\beta>>1$,$(m^\ast>>1)$ there are
several ways to find approximate solutions of this equation, all included
in so called class of the improved WKB methods [15]-[20],[23],[24]
(and the references therein). We shall use the method developed by us
in a series of papers [20], [23],[24] which has the advantage to
make a clear distinction between the various contributions to
the free energy: phonons, 1-kink, 2-kink and so on
sectors. The method is based on well known results from the
theory of asymptotic solutions of second order differential
equations depending on a large parameter [25] an is presented in
the Appendix. As we are interested in the leading term in
$m^\ast$ the exact equation (4) is approximated by the
comparison equation (A.9). With $V_{eff}$ replacing the
potential $V$ in the definition (A.8) of the constant $a$ the
first term in the asymptotic expansion of the lowest eigenvalue
of the isolated potential well is given by
$$E_{0}~=~{1\over 2\sqrt{m^\ast}}\left(1-{l^{2}\over 24d^{2}}\right)~
+~O({1\over m^\ast}).\eqno(6)$$
Then the corresponding contribution to the free energy density
(5) writes
$$F_{0}~=~{1\over \beta l}~ln(\beta h\omega_{0}{d\over l})~+~
{1\over 2\beta d}\left(1-{l^{2}\over 24d^{2}}\right)\eqno(7)$$
and is easily identified with the first terms in the series
expansion in powers of $l^{2}/d^{2}$ of the exact free energy of
an independent gas of phonons with the dispersion relation
$$\omega^{2}(k)~=~\omega_{0}^{2}\left(1+4{d^{2}\over l^{2}}~
sin^{2}~{kl\over 2}\right).\eqno(8)$$
Higher order terms in $(m^\ast)^{-1/2}$, neglected in (6), will
correspond to anharmonic interactions between the phonons [23].

 The presence of the other degenerate minima determines a
symmetrical splitting of each eigenvalue of the isolated
potential well. In the case of the $\phi^4$ model we obtain
$~~E_{0}~\rightarrow~\tilde\epsilon_{0}~=~E_{0}\pm t_{0}$ where
$$t_{0}~=~{c\over \sqrt{m^\ast}}\left(1-{l^{2}\over 24d^{2}}\right)
\nu\eqno(9)$$
with $c=1$. As is explained in the Appendix the value of $\nu$
is determined from the boundary conditions (A.11) imposed on the
eigenfunctions of the comparison equation. For the $\phi^4$
model we get
$$-1~=~2\sqrt{\pi e}~{e^{-i\pi\nu}\over \Gamma(-\nu)}~
exp\left(2\lambda J+\nu(1+ln2)-{1\over 2}(1+2\nu)ln(1+2\nu)\right).
\eqno(10)$$
This expression is valid in leading order in $(m^\ast)^{-1/2}$, but
in any order in $\nu$, and is the starting point to find
multi-kink contributions to the free energy. Here $J$ is an
integral defined by
$$J~=~\int_{0}^{\mu_{1}}\sqrt{V_{eff}-\tilde\epsilon}~d\phi\eqno(11)$$
with $\mu_{1}$ the left turning point of the isolated potential
well. As we are interested in the ground state the quantity
$\tilde\epsilon$ from (11) is connected to $\nu$ by
$$\tilde\epsilon~=~{1\over 2\sqrt{m^\ast}}\left(1-{l^{2}\over 24d^{2}}\right)
(1+2\nu c)~=~\tilde\epsilon_{0}\left(1-{l^{2}\over 24d^{2}}\right)\eqno(12)$$
where $\tilde\epsilon_{0}$ is now independent on the lattice spacing
$l$. It is seen that $J$ depends on $l^{2}/d^{2}$ both through
the integrand and the upper limit of integration. Using (12) and
the expression of $V_{eff}$ the turning points of the $\phi^{4}$
model, calculated in the order $O(l^{2}/d^{2})$, are given by
$$\mu^{2}_{1,2}~=~1\mp\sqrt{8\tilde\epsilon_{0}}+{l^{2}\over 24d^{2}}
\sqrt{8\tilde\epsilon_{0}}\left(\mp{1\over 2}+\sqrt{8\tilde\epsilon_{0}}\right)
+O({l^{4}\over d^{4}}).\eqno(13)$$
In the same order
$$J~=~{1\over 2\sqrt{2}}\int_{0}^{\mu_{1}}\left(1-{l^{2}\over 24d^{2}}
\phi^{2}\right)\sqrt{(\mu_{1}^{2}-\phi^{2})(\mu_{2}^{2}-\phi^{2})}
{}~d\phi~+~O(l^{4}/d^{4}).\eqno(14)$$
Writing
$$J~=~J_{0}~+~{l^{2}\over 24d^{2}}~J_{1}\eqno(15)$$
where now $J_{0}$ and $J_{1}$ doesn't depend any more on
$l^{2}/d^{2}$, we get
$$J_{0}~=~{1\over 2\sqrt{2}}\int_{0}^{\mu_{1}}\sqrt{(\mu^{2}_{1}-
\phi^{2})(\mu^{2}_{2}-\phi^{2})}~d\phi$$
$$J_{1}~\simeq~-{1\over 2\sqrt{2}}\int_{0}^{\mu_{1}}\phi^{2}
\sqrt{(\mu_{1}^{2}-\phi^{2})(\mu_{2}^{2}-\phi^{2})}~d\phi~+$$
$$+~{8\tilde\epsilon_{0}\over 4\sqrt{2}}\int_{0}^{\mu_{1}}
{1-2\phi^{2}\over \sqrt{(\mu_{1}^{2}-\phi^{2})(\mu_{2}^{2}-\phi^{2})}}
{}~d\phi .$$
These integrals can be calculated in terms of complete elliptic
integrals of modulus in the vicinity of unity. In the low
temperature limit, keaping terms up to the order $\tilde\epsilon_{0}$
we find
$$2\sqrt{2}J_0~\simeq~{2\over 3}-\tilde\epsilon_0(1+ln{8\over
\tilde\epsilon_0})
$$
$$2\sqrt{2}J_{1}~\simeq~-{2\over 15}+5\tilde\epsilon_{0}.\eqno(16)$$
Introducing these results into (10) we get
$$\sqrt{{6\beta E_{k}^{(0)}\over\pi}}~e^{-\beta E_{k}^{(0)}(1-{l^{2}
\over 120d^{2}})}~e^{-{5l^{2}\over 48d^{2}}}~=$$
$$=~-~{e^{-i\pi\nu}\over\Gamma(-\nu)}~e^{-\nu~ln(12\beta E^{(0)}_{k}
)}~e^{\nu{5l^{2}\over 24d^{2}}}\eqno(17)-\phi^{4}$$
where $E_{k}^{(0)}={2\over 3}~Ac_{0}\omega_{0}$ is the energy of
the static kink.

 In the sine-Gordon case (periodic potential)
each eigenvalue of the isolated potential well is symmetrically splitted in an
allowed band of width $2t_0$ where $t_0$ is given by (9) with
$c=2$, and $\nu$ results from boundary conditions imposed on the
eigenfunction at the point $\phi=\pi$. Finally we get
$$-1~=~2\sqrt{\pi e}~{e^{i\pi\nu}\over \Gamma(-\nu)}~
{\Gamma({1\over 2}+\nu)\over\Gamma({1\over 2})}~
exp\left(2\lambda J+2\nu (1+ln2)-{1+4\nu\over 2}~ln(1+4\nu)\right),
\eqno(10a)$$
where the integral $J$ is defined by
$$J~=~\int^{\pi}_{\bar{\phi}}~\sqrt{V_{eff}-\tilde\epsilon}~d\phi.\eqno(11a)$$
The integral $J$ is calculated in the same liniar approximation in ${l^{2}
\over d^{2}}$ and the leading terms in $\tilde\epsilon_{0}$ is given by
$$J~=~2\sqrt{2}\left((1-{l^{2}\over 72d^{2}})-{\tilde\epsilon_{0}\over
8}~ln{32\over\tilde\epsilon_0}-{\tilde\epsilon_{0}\over 8}~(1-{l^{2}\over 24d
^{2}})\right)\eqno(16a)$$
where $\tilde\epsilon_{0}$ is defined in (12) being independent on
lattice corrections. Introducing (16a) into (10a) and using the
expression $E_{k}^{(0)}=8Ac_{0}\omega_{0}$ for the energy of the
static kink one obtains
$$\sqrt{{2\beta E_{k}^{(0)}\over \pi}}~e^{-\beta E_{k}^{(0)}(1-
{l^{2}\over 72d^{2}})}~e^{-{l^{2}\over48d^{2}}}~=$$
$$=~-~{e^{i\pi\nu}\over\Gamma(-\nu)}~{\Gamma({1\over2}+\nu)\over
\Gamma({1\over2})}~e^{-2\nu~ln2\beta E_{k}^{(0)}}~e^{2\nu{l^{2}
\over24d^{2}}}.\eqno(17)-SG$$

In order to find various kink contributions we have to expand
the right hand side of equations (17) in powers of $\nu$ and to
write also
$$\nu~=~\nu_{k}~+~\nu_{kk}~+...\eqno(18)$$
The single kink term $\nu_{k}$ comes from the expansion of
$${1\over \Gamma(-\nu)}~=~-\nu+\gamma\nu^{2}$$
($\gamma=0.577$ - Euler's constant) and one obtains easily
$$\nu_{k}~=~\sqrt{{6\beta E_{k}^{(0)}\over\pi}}~e^{-\beta E_{k}^{(0)}
(1-{l^{2}\over120d^{2}})}~\left(1-{5l^{2}\over48d^{2}}\right)\eqno(19)
-\phi^{4}$$
$$\nu_{k}~=~\sqrt{{2\beta E_{k}^{(0)}\over\pi}}~e^{-\beta E_{k}^{(0)}
(1-{l^{2}\over72d^{2}})}\left(1-{l^{2}\over48d^{2}}\right)\eqno(19)-SG$$
This result is in complete agreement with that found by
Trullinger and Sasaki [21].

In finding the kink-kink contribution $\nu_{kk}$ we have to take
into account that the expansion (18) is a formal expansion in
powers of $e^{-\beta E_{k}^{(0)}}$ and consequently $\nu_{kk}$ is
of the same order as $\nu_{k}^{2}$. Then it is easily found that
$$\nu_{kk}~=~\nu^{2}_{k}~\left(\gamma+ln(12\beta E_{k}^{(0)})
-{5l^{2}\over 24d^{2}}+i\pi\right)\eqno(20)-\phi^4$$
One sees that $\nu_{kk}$ gets an imaginary part satisfying the
very simple relation
$$Im~\nu_{kk}~=~\pi\nu^{2}_{k}\eqno(21)$$
This relation has been obtained by Zinn-Justin [26] in his
analysis of the multi-instanton contributions in quantum
mechanics and is tightly related to the non-Borel summability of
the Rayleigh-Schrodinger perturbation expansion when the
potential has degenerate minima.

For the sine-Gordon model one obtains in a similar way
$$\nu_{kk}~=~2\nu^{2}_{k}~(\gamma+ln~4\beta E_{k}^{(0)}-
{l^{2}\over24d^{2}}-{i\pi\over2})\eqno(20)-SG$$
Both $(19-\phi^4)$ and $(20-SG)$ in the limit
${l^{2}\over d^{2}}\rightarrow 0$ are in complete agreement with previous
calculations of the kink-kink sector contributions [26]-[28].

\vskip 2truecm
{\bf 3. Concluding Remarks}
\vskip 1truecm

According to CKBT phenomenology the thermodynamic properties of
the systems described by the Hamiltonians of the form (1) are
influenced by the existence of the static kinks [15]-[19].
A complete agreement between this phenomenology and the exact
results of TIO in the low temperature limit is found if one
takes into account the scattering of phonons on the static kink,
leading to a renormalisation of the kink energy.

The results of Trullinger and Sasaki and of the present paper
are showing that lattice corrections are easily included into the
thermodynamics of this systems. Although a complete proof of a
similar CKBT phenomenology doesn't exist at present, the
existing results strongly support the idea that their
phenomenology is still valid at low temperatures. This comes
both from the phonon part, which, as mentioned above, reproduces
exactly the first terms in the series expansion in ${l^2\over d^2}$
of the free energy of a phonon lattice gas, and from the kink
contribution. As is seen from (19) in the kink
contribution appear a lattice corrected kink energy
$$E_k~=~E_k^{(0)}~-~{l^2\over 12d^2}~E_k^{(1)}\eqno(21)$$
where $E_k^{(0)}$ is the known unperturbed static kink energy
and the correction $E_k^{(1)}$ for the SG and $\phi^4$ model are
given by
$$E_k^{(1)}~=~-{1\over6}~E_k^{(0)},~~~~~~E_k^{(0)}=8A\omega_0 c_0
\eqno(22)-SG$$
$$E_k^{(1)}~=~-{1\over10}~E_k^{(0)},~~~~~E_k^{(0)}={2\over3}A0\omega_0
c_0 \eqno(22)-\phi^4$$
These corrected values have been obtained also by other authors
[5] [6], and  as will be shown below, result from a very simple
perturbation theory. Indeed expanding the field variable
$\phi(x\pm l)$ in a Taylor series the discrete Euler- Lagrange
equation transforms into a fourth order differential equation
$${\partial^{2}\phi\over\partial t^{2}}~+~{l^{2}\over12d^{2}}~
{\partial^{4}\phi\over\partial y^{4}}~+~{\partial^{2}\phi\over
\partial y^{2}}~=~V_{\phi}\eqno(23)$$
where $y=x/d$ and $V_{\phi}={dV\over d\phi}$. Looking for a
perturbed kink solution
$$\phi_{k}(y)~=~\phi_{k}^{(0)}(y)~+~{l^{2}\over12d^{2}}~\phi^{(1)}_{k}(y)
,\eqno(24)$$
where $\phi_k^{(0)}$ is the kink solution of the continuum
limit, the linearized equation determining $\phi_k^{(1)}$ is
$${d^{2}\phi^{(1)}_{k}\over dy^{2}}~-~V_{\phi\phi}(\phi^{(0)}_{k})
{}~\phi_{k}^{(1)}~=~-{d^{4}\phi_{k}^{(0)}\over dy^{4}}.\eqno(25)$$

The homogeneous part of (25) has appeared some years ago in the
study of the kink dynamics in the presence of perturbations and
has as solution the ''translation mode'' (Goldstone mode)~[15],
[16],[29]. Then eq. (25) can be solved by the method of
variation of constants and the obtained results are in complete
agreement with those already existing in literature
[4]-[7],[10]. Using the same Taylor expansion for the field variable one can
write down a corrected integral of energy. Introducing in it the
new expansion of $\phi_k$ (24) the integration is easily
performed and the result is just that presented in (22). Still
unsolved remains the problem of calculation of renormalization of kinks due
to the scattering of lattice phonons on the perturbed kinks,
renormalization which determines the factor multiplying the
exponentials in (19).
\vskip 2 truecm

{\bf Acknoledgement} {\it The authors would like to thank Dr.
Marianne Croitoru for a fruitful colaboration during many years.}

\vfil\eject

{\bf Appendix}
\vskip 1truecm

The lowest eigenvalues of (4) can be found, in the low
temperature limit $m^{\ast}>>1$, using asymptotic methods known
from the theory of second order differential equation depending
on a large parameter [25]. The equation under investigation is
of the type
$$-{1\over 2m^{\star}}{d^{2}\Psi\over d\phi^{2}}~+~V(\phi)\Psi~
=~\tilde{\epsilon}~\Psi,\eqno(A.1)$$
where the potential $V(\phi)$ has at least two degenerate minima
and nonvanishing curvature at the minima.

The problem will be solved in two distinct steps. In the first,
based on Langer's transformation, one looks for an uniform valid
expansion of the solution near the minimum. Due to the
peculiarities mentioned above, this transformation is chosen in
such a way to give the harmonic oscillator behaviour as a
leading term. In the second step, the existance of the other
minima of $V(\phi)$ are taken into account using symmetry
properties of the wave functions. Each eigenvalue of an isolated
potential well is slightely splitted. These tunneling terms are
directly related to the kink contribution to the free energy.

Let $\mu_{1}$ and $\mu_{2}$ be the two turning points of an
isolated potential well. One passes from the variables $\Psi$
and $\phi$ to the new ones $R$ and $x$.
$$\Psi~=~\chi^{-{1\over 4}}R\eqno(A.2)$$
$$\zeta(x)~=~\int_{\mu_{1}}^{\phi}\sqrt{\tilde{\epsilon}-V(\phi)}~d\phi
\eqno(A.3)$$
$$\chi~=~{\tilde{\epsilon}-V\over{(\zeta^{'})^{2}}}.\eqno(A.4)$$
Here $\zeta^{'}={d\zeta\over{dx}}$, and $x$ is still an undefined
function of $\phi$. Then (A.1) transforms into
$${d^{2}R\over{dx^{2}}}~+~\lambda^{2}R~=~\delta R\eqno(A.5)$$
where $\delta=-\chi^{-{3\over 4}}{d^{2}(\chi^{-1\over4})\over{d\phi^{2}}}$
and $\lambda^{2}=2m^{\ast}$.
The function $\chi$ is subjected to the requirement to be
regular and not to vanish in the interval of interest. In order
to obtain the harmonic oscillator behaviour as the leading term,
one takes
$$(\zeta^{'})^{2}~=~4a^{2}(1-x^{2}).\eqno(A.6)$$
This is a function with two simple zeroes and $x=-1$ is
associated with the turning point $\phi =\mu_{1}$ and $x=+1$ to $\phi=\mu_{2}$.
 Now it is possible to determine the relation between the new
variable $x$ and the old one $\phi$. Integrating (A.6) we get
$$a(\pi-\cos ^{-1} x~+~x\sqrt{1-x^{2}})~=~\int_{\mu_{1}}^{\phi}
\sqrt{\tilde\epsilon-V}~d\phi\eqno(A.7)$$
where the constant $a$ is given by
$$a~=~{1\over \pi}~\int_{\mu_{1}}^{\mu_{2}}~\sqrt{\tilde\epsilon-V}~d\phi.
\eqno(A.8)$$
It can be easily proved that for all the domain, even in the
asymptotic region, $\delta \sim {1\over\lambda}$, and taking
into account that $\lambda^{2}=2m^{\ast}>>1$, the first order
approximation can be found from the following comparison
equation [25]:
$${d^{2}R\over dx^{2}}+4a^{2}\lambda^{2}(1-x^{2})R~=~0,\eqno(A.9)$$
whose solutions are the functions of the parabolic cylinder. For
the $\phi^{4}$ model, the only convenient solution ( decreasing
exponentially when $x\rightarrow \infty $ ) is Whittaker's
function $D_{\nu}(y)$, where
$$y=2\sqrt{a\lambda}~x~~~~and~~~\nu=a\lambda-{1\over 2}.\eqno(A.10)$$
The eigenvalues of the isolated potential well are obtained for
$\nu=n$ an integer number.In leading order in $\lambda$ one finds
$$\tilde \epsilon_{n}^{(0)}~\approx~{1\over\sqrt{m^{\ast}}}({1\over2}+n)~
+~O(\lambda^{2}).\eqno(A.11)$$
Now, due to existence of the other degenerate minima of the
potential $V(\phi)$ each eigenvalue of the isolated well is
splitted into two very near levels. They can be found using the
boundary conditions for the wave function and its derivative
$${d\Psi\over d\phi}\vert_{\phi=0}~=~0~,~~~~~~~\Psi\vert_{\phi=0}.
\eqno(A.12)$$
The solution $\Psi(\phi)$ of (A.1) is related to the solution
$R(x)$ of (A.9) by
$$\Psi(\phi)~=~\vert {dx\over d\phi}\vert^{-{1\over 2}}~R(x)~,$$
and the explicit connection formula between the old variable
$\phi$ and the new variable $x$ is (asymptotic region of $x$)
$$x^{2}\simeq ~{1\over a}~\int_{0}^{\mu_{1}}~
\sqrt{\vert\tilde{\epsilon}-V\vert}~d\phi~+~{1\over 2}~+~\ln
2\vert x\vert~+~O(x^{-2}).\eqno(A.13)$$
In applying the boundary conditions (A.11) we have to use the
asymptotic expansion of $D_{\nu}(y)$ [30]
$$D_{\nu}(y)~\sim~y^{\nu}e^{-{y^{2}\over4}}~
-~{\sqrt{2\pi}\over\Gamma(-\nu)}~e^{i\pi\nu}~{1\over y^{1+\nu}}~
e^{{y^{2}\over4}}.\eqno(A.14)$$
The value of $\nu$ is found from a matching between the dominant
$e^{{y^{2}\over4}}$ and the subdominant term $e^{-{y^{2}\over4}}$.
Then from (A.12,A.13,A.14) one obtains
$$-1~=~2\sqrt{\pi e}~{e^{-i\pi\nu}\over\Gamma(-\nu)}~
exp\left(2\lambda~\int_{0}^{\mu_{1}}~\sqrt{V-\tilde{\epsilon}}~d\phi~
+~\nu(1+\ln 2)~-~{1\over2}(1+2\nu)\ln(1+2\nu)\right).\eqno(A.15)$$
The sine-Gordon case can be treated in a very
similar way, the solution for the comparison equation (A.9) being, in that
case,expressed more convenient in terms of Kummer's functions
having a definite parity. Each eigenvalue of an isolated
potential well is splitted into a narrow allowed band. If the
potential $V(\phi)$ has a $2\pi$-periodicity the index $\nu$
corresponding to the lower and upper boundary of the lowest band
follows from similar conditions as (A.12) calculated in the
point $\phi=\pi$. Finally a very similar relation with (A.15) is
found, namely
$$-1=2\sqrt{\pi e}~{e^{-i\pi\nu}\over\Gamma(-{1\over2})}~
{\Gamma({1\over2}+\nu)\over\Gamma({1\over2})}$$
$$exp\left(2\lambda
\int_{\bar\phi}^\pi \sqrt{V-\tilde\epsilon}~d\phi~+~2\nu(1+ln2)-
{1\over2}(1+4\nu)ln(1+4\nu)\right).\eqno(A.15a)$$
Here by $\pm\bar\phi$ we have denoted the turning points of the
isolated potential well, centred arround the minimum $\phi=0$.
Also the definition of $\nu$ is slightly modified and insted of
(A.10) we have $2\nu=a\lambda-{1\over2}$.
To conclude we have to indicate the connection relation between
the symmetric splitting of each eigenvalue of the isolated
potential well, $E_0\rightarrow\tilde\epsilon_0=E_0\pm t_0$ and
the quantity $\nu$. We get
$$t_0~=~{c\over\sqrt{m^\ast}}\nu\eqno(A.16)$$
where $c=1~or~2$ for the $\phi^4$ or the SG model respectively.

\vfil\eject

{\bf References}
\vskip 1truecm
{}~1.~~P. Bak,{\it Rep. Progr. Phys.}{\bf45} (1982) 587  and ref. therein.

{}~2.~~S. Aubry, in{\it ''Solitons and Condensed Matter''}, ed. A. Bishop,

{}~~~~~T. Schneider, {\it Solid St. Sci.}{\bf8} (1978) 264  (Springer,Berlin)

{}~3.~~J.F.Currie, S.E.Trullinger, A.R.Bishop, J.A.Krumhansl,
{\it Phys. Rev.} {\bf B15} (1977)

{}~~~~~ 5567

{}~4.~~J. Andrew Combs, Sidney Yip,{\it Phys. Rev.} {\bf B 28}
{}~(1983) 6873

{}~5.~~Y. Ishimori, T. Munakata,{\it J. Phys. Soc. Japan} {\bf 51}
 (1982) 3367

{}~6.~~C. Willis, M. El-Batanouny, P. Stancioff,{\it Phys. Rev.}
{\bf B 33} (1986) 1904

{}~~~~~P. Stancioff, C. Willis, M. El-Batanouny, S. Burdick, {\it Phys.
Rev.} {\bf B 33} (1986)

{}~~~~~1912

{}~7.~~N. Theodorakopoulos, W. Wunderlich, R. Klein, {\it Solid St.
Comm.} {\bf33} (1980) 213

{}~8.~~M. Peyrard, M.D. Kruskal, {\it Physica }{\bf 13 D} (1984) 88

{}~9.~~M. Peyrard, St. Pnevmaticos, N. Flytzanis, {\it Physica }{\bf 19
D} (1986) 268

{}~~~~~N. Flytzanis, St. Pnevmaticos, M. Peyrard, {\it J. Phys.A: Math.
Gen.}{\bf 22} (1989) 783

10.~~S. De Lillo, {\it Nuovo Cimento }{\bf 100 B} (1987) 105

11.~~M.J. Ablowitz, H. Segur, {\it ''Solitons and Inverse Scattering
Transform''}

{}~~~~~(SIAM Stud. Appl. Math.,Phil. 1981)

12.~~M.J. Ablowitz, J. Ladik, {\it J. Math. Phys. }{\bf 16} (1975)
598;{\bf 17} (1976) 1011

13.~~M. Toda, {\it ''Theory of Nonlinear Lattices'', Solid St. Sci.}
{\bf 20} (Springer, Berlin)

{}~~~~~(1988)

14.~~J,A. Krumhansl, J.R. Schrieffer, {\it Phys. Rev.}{\bf B11} (1975)

{}~~~~~3535

15.~~J.F. Currie, J.A. Krumhansl, A.R. Bishop, S.E.Trullinger,
{\it Phys. Rev. }{\bf B 22}

{}~~~~~(1980)~477

16.~~A.R. Bishop, J.A. Krumhansl, S.E. Trullinger, {\it Physica }{\bf
1 D} (1980) 1

17.~~R.A. Guyer, M.D. Miller, {\it Phys. Rev.} {\bf A17} (1979) 1205

18.~~T. Schneider, E. Stoll, {\it Phys. Rev.} {\bf B22} (1980) 5317

19.~~R.M. De Leonardis, S.E. Trullinger, {\it Phys. Rev. }{\bf B 22}
 (1980) 4558

20.~~M. Croitoru, D. Grecu, A. Visinescu, V. Cionga, {\it Rev.
Roum. Phys.}{\bf 22} (1984)

{}~~~~~853

21.~~S.E. Trullinger, K. Sasaki, {\it Physica} {\bf 28 D},181 (1987)

22.~~G.H. Weiss, A.A. Maradudin, {\it J. Math. Phys. }{\bf 3} (1962) 771

23.~~D. Grecu, A. Visinescu, {\it Physica} {\bf 44 D}, 605 (1990)

{}~~~~~~~~~~~~~~~~~~~~~~~~~~~~~~~~~~~~{\it Rev. Roum.Phys.}{\bf 35}, 45 (1990)

24.~~D. Grecu, A. Visinescu, {\it Rev. Roum. Phys.} {\bf 35}, 3 (1990)

25.~~Ali Hasan Nayfeh, {\it ''Perturbation Methods''} (John Wiley, New
York, 1973)

26.~~J. Zinn-Justin, {\it Nuclear Phys. }{\bf B 192} (1981) 125;{\bf B
218} (1983) 333

{}~~~~~~~~~~~~~~~~~~~~~~{\it J. Math. Phys. }{\bf 22} (1981) 511;~{\bf
25} (1984) 549

27.~~T. Miyashita, K. Maki, {\it Phys. Rev. }{\bf B 23} (1983) 6733;~{\bf B 31}
 (1985) 1836

28.~~K. Sasaki, {\it Progr. Theor. Phys. }{\bf 64} (1982) 411;~{\bf
70} (1983) 593;

{}~~~~~ {\bf 71} (1984) 1169;~{\it Phys. Rev. }{\bf B 33} (1986) 7743

29.~~M.B. Fogel, S.E. Trullinger, A.R. Bishop, J.A. Krumhansl,
{\it Phys. Rev. }{\bf B 15}

{}~~~~~ (1977) 1578

30.~~M. Abramowitz, I.A. Stegun, {\it Handbook of Mathematical Functions,
Nat. Bur. of

{}~~~~~Standards, 1965}

{}~~~~~~M. Magnus, F. Oberhettinger, {\it Formeln und Satze fur
die speziellen

{}~~~~ Functionen der Mathematischen Physik, Springer Verlag,Berlin,1943}

\bye